\documentstyle[prd,aps,epsf,psfig,12pt]{revtex}

\def\beq{\begin{equation}}
\def\eeq{\end{equation}}
\def\ber{\begin{eqnarray}}
\def\eer{\end{eqnarray}}

\def\lsim{\
  \lower-1.5pt\vbox{\hbox{\rlap{$<$}\lower5.3pt\vbox{\hbox{$\sim$}}}}\ }
\def\gsim{\
  \lower-1.5pt\vbox{\hbox{\rlap{$>$}\lower5.3pt\vbox{\hbox{$\sim$}}}}\ }
\def\apj{{Astroph.\@ J.\ }}

\def\prd{{Phys.\@ Rev.\@ D\ }}
\def\pd{{Phys.\@ Rev.\@ D\ }}

\def\etal{{\it et al.}}
\def\ie {{i.e.}}

\def \lleq {\lower0.9ex\hbox{ $\buildrel < \over \sim$} ~}
\def \ggeq {\lower0.9ex\hbox{ $\buildrel > \over \sim$} ~}

\def\lsim{\
  \lower-1.5pt\vbox{\hbox{\rlap{$<$}\lower5.3pt\vbox{\hbox{$\sim$}}}}\ }
  \def\gsim{\
    \lower-1.5pt\vbox{\hbox{\rlap{$>$}\lower5.3pt\vbox{\hbox{$\sim$}}}}\ }

\begin{document}

\tightenlines

\title{Did the universe loiter at high redshifts?}

\author{Varun Sahni$^a$ and Yuri Shtanov$^b$}
\address{$^a$Inter-University Centre for Astronomy and Astrophysics, Post Bag 4,
Ganeshkhind, Pune 411~007, India \\
$^b$Bogolyubov Institute for Theoretical Physics, Kiev 03143, Ukraine}

\maketitle

\bigskip

\begin{abstract}
We show that loitering at high redshifts ($z \gsim 6$) can easily arise in
braneworld models of dark energy which, in addition to being spatially flat,
also accelerate at late times. Loitering is characterized by the fact that the
Hubble parameter dips in value over a narrow redshift range which we shall
refer to as the `loitering epoch'. During loitering, density perturbations are
expected to grow rapidly. In addition, since the expansion of the universe
slows down, its age near loitering dramatically increases. An early epoch of
loitering is expected to boost the formation of high redshift gravitationally
bound systems such as $10^9 M_\odot$ black holes at $z \sim 6$ and lower-mass
black holes and/or Population III stars at $z > 10$, whose existence could be
problematic within the LCDM scenario.
Loitering models also help to reduce the redshift of reionization from its
currently (high) value of $z_{\rm reion} \simeq 17$ in LCDM cosmology, thus
alleviating a significant source of tension between observations of the
high-redshift universe and theoretical model building.
Currently a loitering universe accelerates with an effective equation of state
$w < -1$ thus mimicking phantom dark energy. Unlike phantom, however, the late-time
expansion of the universe in our model is singularity free, and a universe
that loitered in the past will approach a LCDM model asymptotically in the
distant future.

\end{abstract}


\vfill\eject
\section{Introduction}

Standard Big Bang cosmology, as epitomized in its most recent avatar, LCDM, is
in excellent agreement with a host of cosmological observations including
galaxy clustering, fluctuations in the CMB, and
the current accelerating epoch. Yet it appears that recent observations at
modest redshifts ($6\lsim z \lsim 20$) may have some surprises in store for
LCDM.

\begin{itemize}

\item[(i)]  In less than a decade of observations,
the number of known high redshift QSO's has shown an almost twenty-fold
increase\,!
Indeed, over $400$ QSO's with redshifts $z > 4$ are known at present, and the
seven highest redshift quasars have $z > 5.7$ \cite{richards03}. If quasars
shine by virtue of an accreting black hole at their centers, then all these
QSO's must host $\gsim 10^9 M_\odot$ black holes. Whether such highly massive
black holes can successfully form in a LCDM universe which is less than a
billion years old at $z \sim 6$ remains an open question, but most theorists
seem to agree that theoretical models of the growth of black holes, whether by
accretion or through BH--BH mergers, are under some tension to explain the
observations \cite{richards03,bh}.

\item[(ii)]  In addition to the presence of large supermassive black holes at
$z \sim 6$, there is indirect evidence to suggest that a population of less
massive black holes and/or Population~III stars was already in place by $z
\gsim 17$ and may have been responsible for ionizing the universe at lower
redshifts.\footnote{WMAP observations give $\tau = 0.17 \pm 0.06$ for the
optical depth which translates into $z = 17 \pm 5$ for the reionization
redshift in LCDM cosmology \cite{wmap}.} Whether the LCDM model can form
structure early and efficiently enough to successfully reionize the universe by
$z \sim 17$ is a moot point \cite{reionization}.
In any case, both (i) and (ii) provoke the concerned
cosmologist to look for
alternative models, which, while preserving the manifold strengths and
successes of LCDM, will also be able to provide a compelling resolution to the
issues raised above. In this paper, we show that one such model --- a
braneworld universe which loiters at an early epoch --- may provide an
attractive alternative to LCDM.

\end{itemize}

\section{Loitering universe}

A considerable body of evidence exists to suggest that the universe is
currently accelerating, \ie, that its expansion rate is speeding up rather than
slowing down \cite{sn}. Models of dark energy incorporate this effect by making
the deceleration parameter change sign while the Hubble parameter is usually
assumed to be a monotonically decreasing function of the cosmic
time.\footnote{Phantom models may provide an exception to this rule, see
\cite{sahni04} and references therein.} In the present paper, we show that this
need not necessarily be the case and that compelling dark energy models can be
constructed in which $H(z)$ dips in value at high redshifts. In these models,
$dH(z)/dz \simeq 0$ at $z_{\rm loit} \gg 1$, which is called the `loitering
redshift'. (A universe which loiters has also been called a `hesitating'
universe, since, if $H(z_{\rm loit}) \simeq 0$, the universe hesitates at the
redshift $z_{\rm loit}$ for a lengthy period of time --- before either
collapsing or re-expanding.) Loitering increases the age of the universe at
high $z$ and also provides a boost to the growth of density inhomogeneities,
thereby endowing a dark energy model with compelling new properties.

In this paper, we show that loitering can arise naturally in a class of
braneworld models which also provide a viable alternative to LCDM in explaining
the late-time acceleration of the universe \cite{SS}. Before we discuss
loitering in braneworld models, let us briefly review the status of loitering
in standard General Relativity. Within a FRW setting, loitering can only arise
in a universe which is spatially closed and which is filled with matter and a
cosmological constant (or some other form of dark energy --- see \cite{sfs92}).
The evolution of such a universe is described by the equation
\beq
H^2 = \frac{8\pi G}{3}\frac{\rho_{0} a_0^3}{a^3} + \frac{\Lambda}{3} -
\frac{\kappa}{a^2} \, , \quad \kappa = 1 \, , \label{eq:frw}
\eeq
where $\rho_{0}$ is the present matter density. Loitering in (\ref{eq:frw})
arises if the curvature term ($1/a^2$) is large enough to substantially offset
the dark-matter + dark-energy terms but not so large that the universe
collapses. The redshift at which the universe loitered can be determined by
rewriting (\ref{eq:frw}) in the form
\beq
h^2(z) \equiv {H^2(z) \over H_0^2} =
\Omega_{\rm m} (1\!+\!z)^3  + \Omega_{\Lambda} +
\Omega_{\kappa} (1\!+\!z)^2  \, , \label{eq:frw1}
\eeq
where $\Omega_\kappa = -\kappa/a_0^2H_0^2$, $\Omega_{\rm m} = 8\pi
G\rho_{0}/3H_0^2$, $\Omega_{\Lambda} = \Lambda/3H_0^2$,
the subscript `{\small 0}' refers
to present epoch, and the constraint equation
requires $\Omega_\kappa = 1 - \Omega_{\rm m} - \Omega_{\Lambda}$.
The loitering condition $dh/dz = 0$ gives
\beq
1 + z_{\rm loit} = \frac{2\vert\Omega_\kappa\vert}{3 \Omega_{\rm m}} \, ,
\eeq
and it is easy to show that $z_{\rm loit} \leq 2$ for $\Omega_{\rm m} \geq 0.1$
\cite{sfs92}. (Note that a large value of $\vert\Omega_\kappa\vert$ can cause
the universe to recollapse.) The value of the Hubble parameter at loitering can
be determined by substituting $z_{\rm loit}$ into (\ref{eq:frw1}). Note that,
since ${\ddot a}/a = {\dot H} + H^2$, it follows that $({\ddot
a}/a)\big\vert_{z=z_{\rm loit}} = H^2(z_{\rm loit})$ at loitering. (The special
case ${\dot a} = 0$, ${\ddot a} = 0$ corresponds to the static Einstein
universe \cite{ss00}. For a detailed discussion of loitering in FRW models with
dark energy see \cite{sfs92}. Loitering in more general contexts has been
discussed in \cite{robert,polarski}.)

Interest in loitering FRW models has waxed and waned ever since the original
discovery of a loitering cosmology by Lema\^{\i}tre over seventy years ago
\cite{lemaitre}. Among the reasons why the interest in loitering appears to
have declined in more recent times are the following:
%NEW
(i)~even though loitering models can accommodate an accelerating universe,
the loitering redshift is usually small: $z_{\rm loit} \leq 2$ in LCDM;
(ii)~loitering models require a large spatial curvature, which is at variance
with inflationary predictions and CMB observations both of which support a flat
universe.
As we shall show, in marked contrast with the above scenario,
loitering in braneworld models can take place
in a spatially flat universe and at high redshifts ($z \gsim 6$). At late
times, the loitering braneworld model has properties similar to those of LCDM.

\section{Loitering in braneworld models} \label{loitering}

The braneworld model which we shall consider presents a successful synthesis of
the higher-dimensional ansatzes proposed by Randall and Sundrum \cite{RS} and
Dvali, Gabadadze, and Porrati \cite{DGP}, and is described by the action
\cite{CHD}
\ber \label{action}
S &=& M^3  \left[ \int_{\rm bulk} \left({\mathcal R} - 2 \Lambda_{\rm b}
\right) - 2 \int_{\rm brane} K  \right] + \int_{\rm brane} \left( m^2 R -
2 \sigma \right) \nonumber\\
&+& \int_{\rm brane} L (h_{ab}, \phi) \, .
\eer
Here, ${\mathcal R}$ is the scalar curvature of the five-dimensional metric
$g_{ab}$ in the bulk, and $R$ is the scalar curvature of the induced metric
$h_{ab} = g_{ab} - n_a n_b$ on the brane, where $n^a$ is the vector field of
the inner unit normal to the brane. The quantity $K = K_{ab} h^{ab}$ is the
trace of the symmetric tensor of extrinsic curvature $K_{ab} = h^c{}_a \nabla_c
n_b$ of the brane, and $L (h_{ab}, \phi)$ denotes the Lagrangian density of the
four-dimensional matter fields $\phi$ whose dynamics is restricted to the brane
(we use the notation and conventions of \cite{Wald}). Integrations over the
bulk and brane are taken with the natural volume elements $\sqrt{- g}\, d^5 x$
and $\sqrt{- h}\, d^4 x$, respectively. The constants $M$ and $m$ denote,
respectively, the five-dimensional and four-dimensional Planck masses,
$\Lambda_{\rm b}$ is the five-dimensional (bulk) cosmological constant, and
$\sigma$ is the brane tension.

Action (\ref{action}) leads to the following expression for the Hubble
parameter on the brane for a {\em spatially flat\/} universe \cite{SS}:
\begin{equation}\label{hubble}
H^2 (a)  = {A \over a^3} + B + {2 \over \ell^2} \left[ 1 \pm \sqrt{1 + \ell^2
\left({A \over a^3} + B  - {\Lambda_{\rm b} \over 6} - {C \over a^4} \right)}
\right] \, ,
\end{equation}
where
\begin{equation}\label{ab}
A = {\rho_{0} a_0^3 \over 3 m^2} \, , \quad B = {\sigma \over 3 m^2} \, , \quad
\ell = {2 m^2 \over M^3} \, .
\end{equation}
Note that the four-dimensional Planck mass $m$ is related to the effective
Newton's constant on the brane as $m = 1/\sqrt{8\pi G}$.

The two signs in (\ref{hubble}) correspond to the two branches of the braneworld
models and are connected with the two different ways in which
the brane can be embedded in the bulk. As shown in \cite{SS}, the `$+$' sign in
(\ref{hubble}) corresponds to late time acceleration of the universe driven by
dark energy with an `effective' equation of state $w \geq -1$ (BRANE2) whereas
the `$-$' sign is associated with phantom-like behaviour $w \leq -1$ (BRANE1).
The length scale $\ell = {2 m^2 / M^3} \sim cH_0^{-1}$ in a braneworld which
begins to accelerate at the current epoch \cite{DGP,SS}. In particular, when
$\ell = 0$ (corresponding to $m = 0$), equation (\ref{hubble}) reduces to
\begin{equation}\label{cosmolim}
H^2  = {\Lambda_{\rm b} \over 6} + {C \over a^4} + {\left(\rho + \sigma
\right)^2 \over 9 M^6} \,  ,
\end{equation}
describing the evolution of a RS braneworld  \cite{BDL}.
The opposite limit $\ell \to \infty$ ($M = 0$) results in the LCDM model
\beq
H^2 (a) = {A \over a^3} + B \, ,
\eeq
while setting $\Lambda_{\rm b} = 0$ and $\sigma = 0$ gives rise to the DGP
braneworld \cite{DGP}.

Of crucial importance to the present analysis will be the `dark radiation' term
$C/a^4$  in (\ref{hubble}) whose presence is a generic feature in braneworld
models and which describes the projection of the bulk degrees of freedom onto
the brane. (It corresponds to the presence of the bulk black hole.) An
interesting situation arises when $C< 0$ and $\ell^2 {|C|/a^4} \gg 1$. In this
case, if $\ell^2 {|C|/a^4}$ is larger than the remaining terms under the square
root in (\ref{hubble}), then that equation reduces to\footnote{The negative
value of the dark-radiation term
implies the presence of black hole with negative mass --- hence, naked
singularity --- in the complete extension of the bulk geometry.  In principle,
this singularity could be ``closed from our view'' by another (invisible)
brane.}
\begin{equation}
H^2(a) \approx {A \over a^3} + B \pm {2 \sqrt{- C} \over \ell a^2} \, .
\label{eq:loit1}
\end{equation}
Equation (\ref{eq:loit1}) bears a close formal resemblance to (\ref{eq:frw}),
which gave rise to loitering solutions in standard FRW geometry for $\kappa =
1$. Indeed, the role of the spatial curvature in (\ref{eq:loit1}) is played by
the dark-radiation term; consequently, a spatially open universe is mimicked by
the BRANE2 model while a closed universe is mimicked by BRANE1. In analogy with
standard cosmology, one might expect the braneworld model (\ref{hubble}) to
show loitering behaviour in the BRANE1 case. This is indeed the case, and
stronly loitering solutions to (\ref{hubble}) \& (\ref{eq:loit1}) can be found
by requiring $H'(a)=0$.
\begin{figure*}
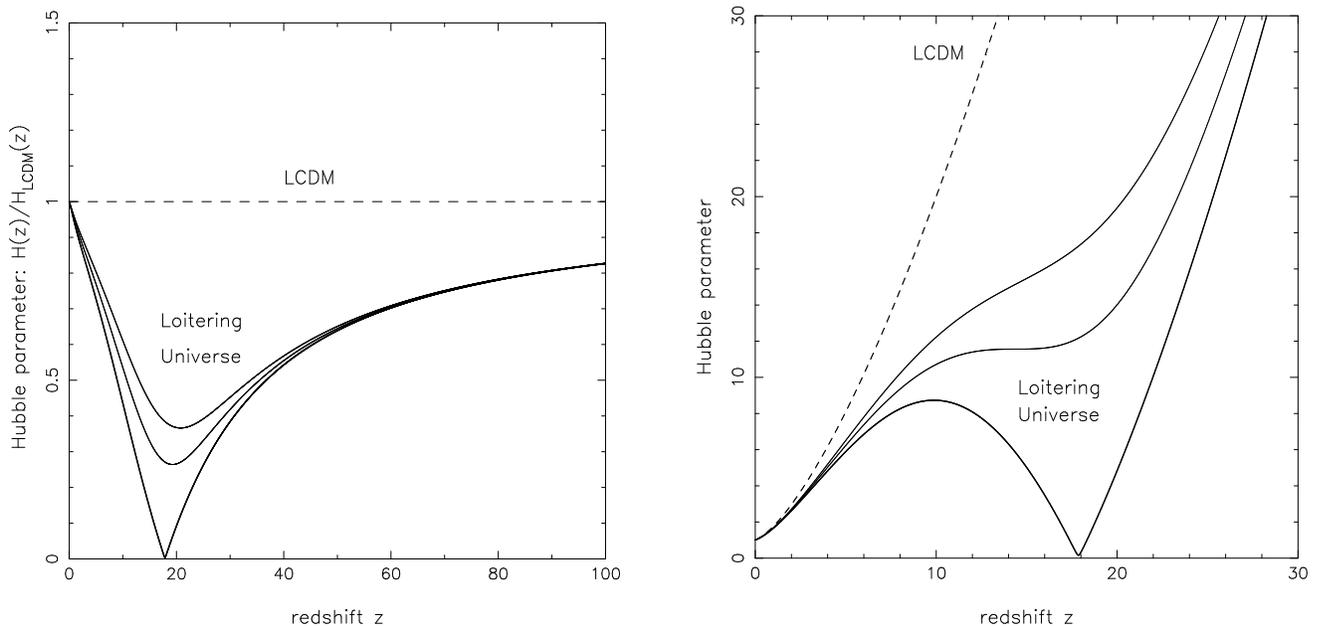

%\centering
\begin{center}
%$\begin{array}{c@{\hspace{0.2in}}c}
$\begin{array}{c@{\hspace{0.4in}}c}
\multicolumn{1}{l}{\mbox{}} &
\multicolumn{1}{l}{\mbox{}} \\ [0.1cm]
\epsfxsize=3.2in
\epsffile{hub_loiter2.epsi} &
\epsfxsize=3.2in
\epsffile{hub_loiter1.epsi} \\ [0.20cm]
%\mbox{\bf (a)} & \mbox{\bf (b)}
\end{array}$
\end{center}
\caption{\small The Hubble parameter for a universe that loiters at $z_{\rm
loit} \simeq 18$. Parameter values are $\Omega_{\rm m} = 0.3$, $\Omega_C =
8.0$, $\Omega_\ell = 3.0$, and $\Omega_{\Lambda_{\rm b}}/10^5 = 6,\, 4.5,\,
3.4$ (solid lines, from top to bottom). The left panel shows the Hubble
parameter with respect to the LCDM value while, in the right panel, the LCDM
(dashed) and loitering (solid)
Hubble parameters are shown separately. } \label{fig:hubble}
\end{figure*}

Although this is the general procedure which we follow, for practical purposes
it will be more suitable to rewrite (\ref{hubble}) with the lower sign in the
form
\ber \label{hubble1}
{H^2(z) \over H_0^2} &=& \Omega_{\rm m} (1\!+\!z)^3 + \Omega_\sigma + 2
\Omega_\ell \nonumber \\ &-& 2 \sqrt{\Omega_\ell}\, \sqrt{\Omega_{\rm m}
(1\!+\!z )^3 + \Omega_\sigma + \Omega_\ell + \Omega_{\Lambda_{\rm b}} +
\Omega_C(1\!+\!z )^4} \, ,
\eer
where
\ber \label{omegas}
\Omega_{\rm m} =  {\rho_0 \over 3 m^2 H_0^2} \, , \quad \Omega_\sigma = {\sigma
\over 3 m^2 H_0^2} \, , \quad \Omega_\ell = {1 \over \ell^2 H_0^2} \, ,
\nonumber \\ \Omega_{\Lambda_{\rm b}} = - {\Lambda_{\rm b} \over 6 H_0^2} \, ,
\quad \Omega_C = -\frac{C}{a_0^4H_0^2}\, .
\eer
The $\Omega_i's$ satisfy the constraint equation
\begin{equation} \label{omega-r1}
\Omega_{\rm m} + \Omega_\sigma - 2 \sqrt{\Omega_\ell}\, \sqrt{1 +
\Omega_{\Lambda_{\rm b}} + \Omega_C} = 1 \, .
\end{equation}

When the dark-radiation term is strongly dominating, Eq.~(\ref{hubble}) reduces
to
\beq
{H^2(z) \over H_0^2} \simeq \Omega_{\rm m}(1\!+\!z)^3 +
\Omega_\sigma - 2 \sqrt{\Omega_\ell \Omega_C}(1\!+\!z)^2 \, ,
\label{eq:hubble2}
\eeq
which is the braneworld analog of (\ref{eq:frw1}). The loitering redshift in
this case can be defined by the condition $H'(z_{\rm loit}) = 0$; as a result, one gets
\beq
1\!+\!z_{\rm loit} \simeq \frac{4}{3} \, \frac{\sqrt{\Omega_C\Omega_\ell}}{\Omega_{\rm
m}} \, . \label{eq:loit}
\eeq
From this expression we find that the universe will loiter at a large redshift
$z_{\rm loit} \gg 1$ provided $\Omega_C \Omega_\ell \gg \Omega_{\rm m}^2$. Since
$\Omega_{\rm m}^2 \ll 1$, this is not difficult to achieve in practice.
Successful loitering of this type requires the following two conditions to
be satisfied:
\ber
&\Omega_C(1\!+\!z_{\rm loit})^4 \gg \Omega_{\rm m}(1\!+\!z_{\rm loit})^3 +
\Omega_\sigma +
\Omega_\ell + \Omega_{\Lambda_{\rm b}} \, , \nonumber\\
&\Omega_\sigma \sim \sqrt{\Omega_\ell \Omega_C}(1\!+\!z_{\rm loit})^2 \, .
\label{eq:constraint}
\eer
The first inequality ensures that the dark-radiation term dominates over the
remaining terms under the square root of (\ref{hubble1}) during loitering,
while the second makes sure that this term is never so large as to cause the
universe to recollapse.

Substituting the value for $1\!+\!z_{\rm loit}$ from (\ref{eq:loit}) into
(\ref{eq:constraint}), we obtain
\beq
\Omega_\sigma \sim \frac{(\Omega_C\Omega_\ell)^{3/2}}{\Omega_{\rm m}^2} \gg
\Omega_\ell \, , \label{eq:constraint1}
\eeq
which is a necessary condition for loitering in our braneworld model.

Finally, the Hubble parameter at loitering is given by the approximate
expression
\beq
{H^2(z_{\rm loit}) \over H_0^2} \simeq \Omega_\sigma -
\frac{32}{27}\frac{(\Omega_C\Omega_\ell)^{3/2}}{\Omega_{\rm m}^2} \, .
\label{eq:hub_loiter}
\eeq

Note that conventional loitering is usually associated with a vanishingly small
value for the Hubble parameter at the loitering redshift \cite{sfs92}. The
Hubble parameter at loitering can be set as close to zero as possible; however,
we do not require it to be {\em very\/} close to zero. A small `dip' in the
value of $H(z)$, which is sufficient for our purposes, arises for a far larger
class of parameter values than the more demanding condition $H(z_{\rm loit})
\simeq 0$.

Moreover, in a wide range of parameters, the universe evolution may
not exhibit a minimum of the Hubble parameter $H(z)$.  In this case, the definition
of the loitering redshift by the condition $H'(z_{\rm loit}) = 0$ is not appropriate
and can be generalised in several different ways, one of which is described in
the appendix.

An example of a loitering model is shown in Fig.~\ref{fig:hubble}, where the
Hubble parameter of a universe which loitered at $z \simeq 18$ is plotted
against the redshift, keeping $\Omega_{\rm m}$, $\Omega_\ell$, and $\Omega_C$
fixed and varying the value of $\Omega_{\Lambda_{\rm b}}$.
The right-hand panel of Fig.~\ref{fig:hubble} illustrates the fact that the
loitering universe can show a variety of interesting behaviour: (i)~top curve,
$H(z)$ is monotonically increasing and $H'(z) \simeq {\rm constant}$ in the
loitering interval; (ii)~middle curve, $H(z)$ appears to have an inflexion
point ($H' \simeq 0$, $H'' \simeq 0$) during loitering; (iii)~lower curve, $H(z)$
has both a maximum and a mininimum, the latter occuring in the loitering
regime.

At this point, we would like to stress an important difference existing between
the Randall--Sundrum braneworld (\ref{cosmolim}) and our universe
(\ref{hubble}) due to which the latter can accommodate a large value of dark
radiation without violating nucleosynthesis constraints whereas the former
cannot. In the Randall--Sundrum braneworld (\ref{cosmolim}), the dark-radiation
term ($C/a^4$) affects cosmological expansion in {\em exactly the same way\/}
as the usual radiation density $\rho_{\rm r}$, so that this model comes into
serious conflict with the predictions of the big-bang nucleosynthesis if $|C|$
is very large \cite{ichiki}. In the loitering braneworld, on the other hand,
the dark-radiation term resides under the square root in (\ref{hubble}); due to
this circumstance its effect on the cosmological expansion is less severe and,
more importantly, {\em transient\/}. Indeed, even if the dark-radiation term is
very large ($|C|/a^4 > \rho_{\rm m},\,\rho_{\rm r}$), its influence on
expansion can only be $\propto 1/a^2$, which does not pose a serious threat to
the standard predictions of the big-bang nucleosynthesis.

\begin{figure*}
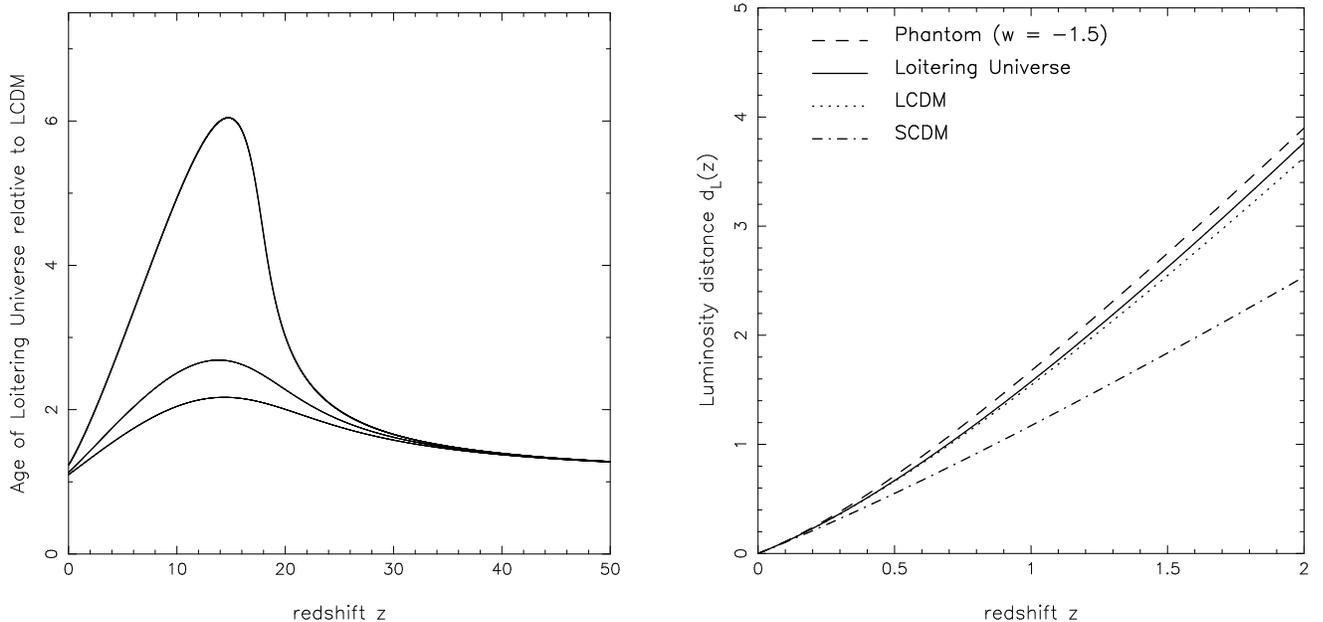

%\centering
\begin{center}
%$\begin{array}{c@{\hspace{0.2in}}c}
$\begin{array}{c@{\hspace{0.4in}}c}
\multicolumn{1}{l}{\mbox{}} &
\multicolumn{1}{l}{\mbox{}} \\ [0.1cm]
\epsfxsize=3.2in
\epsffile{age.epsi} &
\epsfxsize=3.2in
\epsffile{lum_loiter.epsi} \\ [0.20cm]
%\mbox{\bf (a)} & \mbox{\bf (b)}
\end{array}$
\end{center}

%\begin{figure*}
%\centering
%\begin{center}
%\epsfxsize=3.4in
%\epsffile{age.epsi}
%\end{center}

\caption{\small In the left panel,
the age of three loitering models is shown relative to the age
in LCDM (model parameters are the same as in Fig.~\ref{fig:hubble}). Note that
the age of the universe near loitering ($z_{\rm loit} \sim 18$) is
significantly greater than that in LCDM although, at the present epoch, the
difference in ages between the two models is relatively small.
In the right panel,
the luminosity distance in a universe that loiters at $z_{\rm
loit} \simeq 18$ is shown in comparison with other models. Note that the
luminosity distance in the loitering model is only slightly larger than that in
LCDM and smaller than that in a phantom model with $w = -1.5$.}

\label{fig:age}
\end{figure*}

A loitering universe could have several important cosmological consequences:
\begin{itemize}

\item The age of the universe during loitering {\em increases\/}, as shown in
Fig.~\ref{fig:age}. The reason for this can be seen immediately from the
expression
\beq
t(z) = \int_z^\infty \frac{dz'}{(1+z')H(z')} \, . \label{eq:age}
\eeq
Clearly, a lower value of $H(z)$ close to loitering will boost the age of the
universe at that epoch. In Fig.~\ref{fig:age}, the age of the universe is
plotted with reference to a LCDM universe, which has been chosen as our
fiducial model. It is interesting to note that, while the {\em age at
loitering\/} can be significantly larger in the loitering model than in LCDM
$\left[t_{\rm loit}(z_{\rm loit}) \sim {\rm few}\times t_{\rm LCDM}(z_{\rm
loit})\right]$, the present age of the universe in both models is comparable
$\left[t_{\rm loit}(0) \lsim 1.2\times t_{\rm LCDM}(0)\right]$.\footnote{The
age of a LCDM universe at $z \gg 1$ is ~$t(z) \simeq (2/3H_0\sqrt{\Omega_{\rm
m}}) (1+z)^{-3/2} = 5.38 \times 10^8 (1+z/10)^{-3/2} $ years for $\Omega_{\rm
m} = 0.3$ and $h = 0.7$.}
An important consequence of having a larger age of the universe at $z \sim 20$
(or so) is that astrophysical processes at these redshifts have more time in
which to develop. This is especially important for gravitational instability
which forms gravitationally bound systems from the extremely tiny fluctuations
existing at the epoch of last scattering. Thus, an early loitering epoch may be
conducive to the formation of Population~III stars and low-mass black holes at
$z \sim 17$ and also of $\sim 10^9 M_\odot$ black holes at lower redshifts ($z
\sim 6$).

\item In Fig.~\ref{fig:age}, the luminosity distance for the loitering model is
shown, again with LCDM as the base model. The luminosity distance is related to
the Hubble parameter through
\beq
\frac{D_L(z)}{1+z} = \int_0^z\frac{dz}{H(z)} \, . \label{eq:lum_dis}
\eeq
One finds from Fig.~\ref{fig:age} that the luminosity distance in the loitering
model, although somewhat larger than in LCDM, is smaller than in a phantom
model with $w=-1.5$. Since both phantom and LCDM models provide excellent fits
to type~Ia supernova data \cite{sn,caldwell03,alam2}, we expect our family of
`high redshift loitering models' to also be in good agreement with
observations. (A detailed comparison of loitering models with observations
lies outside of the scope of the present paper and will be reported
elsewhere.)

The reason why both the luminosity distance and the current age of the universe
have values which are close to those in the LCDM model is clear from
Fig.~\ref{fig:hubble}, where we see that the difference between the Hubble
parameters for the loitering models and LCDM model is small at low redshifts.
Since both $D_L(z)$ and $t(z)$ probe $H^{-1}(z)$, and since the value of the
Hubble parameter at low $z$ is much smaller than its value at high $z$ (unless
parameter values are chosen to give $H(z_{\rm loit}) \simeq 0$ with a high
precision), it follows that $|D_L^{\rm loit}(z) - D_L^{\rm LCDM}(z)| \ll
D_L^{\rm LCDM}(z)$ and $|t^{\rm loit}(z) - t^{\rm LCDM}(z)| \ll t^{\rm
LCDM}(z)$ for $z \ll 1$.

\item The growth of density perturbations depends sensitively upon the
behaviour of the Hubble parameter, as can be seen from the following equation
describing the growth of linearized density perturbations $\delta =
(\rho-{\bar\rho})/{\bar\rho}$ in a FRW universe (ignoring the effects of pressure):
\beq
{\ddot \delta} + 2H{\dot\delta} - 4\pi G{\bar\rho}~\delta = 0 \, .
\label{eq:delta}
\eeq
In Eq.~(\ref{eq:delta}), the second term $2H{\dot\delta}$ damps the growth of
perturbations; consequently, a lower value of $H(z)$ during loitering will
boost the growth rate in density perturbations, as originally demonstrated in
\cite{sfs92}.

Here we should note that Eq.~(\ref{eq:delta}) for perturbations is perfectly
valid only in general relativity and, in principle, may be corrected or
modified in the braneworld theory under consideration. Thus, for the DGP
braneworld model \cite{DGP} (which corresponds to setting $\sigma = 0$,
$\Lambda_{\rm b} = 0$ and $C =0$ in Eq.~(\ref{hubble})), the linearized
equation
\begin{equation}
{\ddot \delta} + 2H{\dot\delta} - 4\pi G{\bar\rho}\left(1 +
\frac{1}{3\beta}\right) \delta = 0
\label{eq:delta1}
\end{equation}
was derived in \cite{lss}, where
\begin{equation}
\beta = - {1 + \Omega_{\rm m}^2 (t) \over 1 - \Omega_{\rm m}^2 (t)} \, ,
\qquad \Omega_{\rm m} (t) \equiv {8 \pi G \bar \rho (t) \over 3 H^2 (t)} \, .
\end{equation}
It is important to note the similarities as well as differences between
(\ref{eq:delta}) and (\ref{eq:delta1}). Thus, cosmological expansion works in
the same way for both models and introduces the damping term $2H{\dot\delta}$
in (\ref{eq:delta}) as well as in (\ref{eq:delta1}). However, in contrast to
(\ref{eq:delta}), the braneworld perturbation equation (\ref{eq:delta1}) has a
time-dependent (decreasing) effective gravitational constant
\begin{equation}
G_{\rm eff} = G \left(1 + \frac{1}{3\beta}\right) \, ~,
\end{equation}
which is expected to affect the growth rate of linearized density perturbations
in this model.
For the generic braneworld model which we study in this paper
(which has nonzero brane and
bulk cosmological constants and especially nonzero dark radiation:
$C \ne 0$ in Eq.~(\ref{hubble})), the corresponding equation
for cosmological perturbations remains to be derived.
We expect the form of this
equation to be dependent on the additional boundary conditions in the bulk or on the
brane.  However, we anticipate that such an equation will
contain the damping
term $2H{\dot\delta}$ which serves to
enhance the growth of perturbations
%on the spatial scales well below the Hubble scale
in the case of loitering.  At the same time, braneworld-specific effects
%in the
%perturbation
%equation (still to be derived)
may act in the opposite direction leading to the suppression of the growth of
perturbations relative to the FRW model, as is the case, for instance, with the
last term in (\ref{eq:delta1}) for the DGP model \cite{lss}. This is an
important  issue requiring further investigation, and we shall return to it in
a future paper.

\item The deceleration parameter $q$ and the effective equation of state $w$ in our
loitering model are given by the expressions
\ber
q(z) &=& \frac{H'(z)}{H(z)} (1+z) - 1 \, ,\nonumber\\
w(z) &=& {2 q(z) - 1 \over 3 \left[ 1 - \Omega_{\rm m}(z) \right] } \, ,
\eer
where $H(z)$ is determined from (\ref{hubble1}) and
(\ref{omega-r1}). The current values of these quantities are
\ber
q_0 &=& \frac{3}{2}\Omega_{\rm m}\left [1 -
\frac{\sqrt{\Omega_\ell}}{\sqrt{\Omega_\ell} + \sqrt{1 + \Omega_{\Lambda_{\rm
b}} + \Omega_C}}
\left(1 + \frac{4}{3}\frac{\Omega_C}{\Omega_{\rm m}}\right)\right ] - 1 \, ,\nonumber\\
w_0 &=& -1 - \frac{\Omega_{\rm m}}{(1-\Omega_{\rm m})} \cdot
\frac{\sqrt{\Omega_\ell}} {\sqrt{\Omega_\ell} + \sqrt{1 + \Omega_{\Lambda_{\rm
b}} + \Omega_C}} \left(1 + \frac{4}{3}\frac{\Omega_C}{\Omega_{\rm m}} \right)
\, , \label{eq:w_0}
\eer
From Eq.~(\ref{eq:w_0}) we find that $w_0 < -1$ if $\Omega_C \geq 0$; in other
words, our loitering universe has a phantom-like effective equation of state.
%NEW
(In particular, for the loitering models shown in Fig.~\ref{fig:hubble}, we
have $w_0 = -1.035\,$, $-1.04\,$, $-1.047$ (top to bottom), all of which are in
excellent agreement with recent observations \cite{seljak04}.)
However, in contrast to phantom models, the Hubble parameter in a loitering
universe (\ref{hubble1}) does not encounter a future singularity since
$\Omega_C,\,\Omega_\sigma > 0$ is always satisfied in models which loitered in
the past. (Future singularities can arise in braneworld models if $\Omega_C,\,
\Omega_\sigma < 0$ --- see \cite{ss02} for a comprehensive discussion of this
issue and \cite{odintsov} for related ideas.)

An interesting consequence of the loitering braneworld is that
the time-dependent density parameter
$\Omega_{\rm m} (z) = 8\pi G \rho_{\rm m} (z) /3H^2 (z)$
{\em exceeds\/} unity at
some time in the past. This follows immediately from the fact that, since the
value of $H(z)$ in the loitering braneworld model is {\em smaller\/} than its
counterpart in LCDM, the value of $\Omega_{\rm m}(z)$ is larger than its
counterpart in LCDM. One important consequence of this behaviour is that, as
expected from (\ref{eq:w_0}), the effective equation of state (EOS) blows up
precisely when $\Omega_{\rm m} (z) =1$. In Fig.~\ref{fig:state}, we show that, in
contrast to the singular behaviour of the EOS, the deceleration parameter
remains finite and well behaved even as $w \to \infty$. Note that the finite
behaviour of $q(z)$ reflects the fact that the EOS for the braneworld is an
{\em effective\/} quantity and not a real physical property of the theory
--- see \cite{alam} for a related discussion of this issue and
\cite{linder04} for an example of a different dark energy model displaying similar
behaviour. (The deceleration
parameter experiences near-singular behaviour at the higher, loitering
redshift, as $H \to 0$ so that $q \to \infty$.)

\begin{figure*}
\centerline{ \psfig{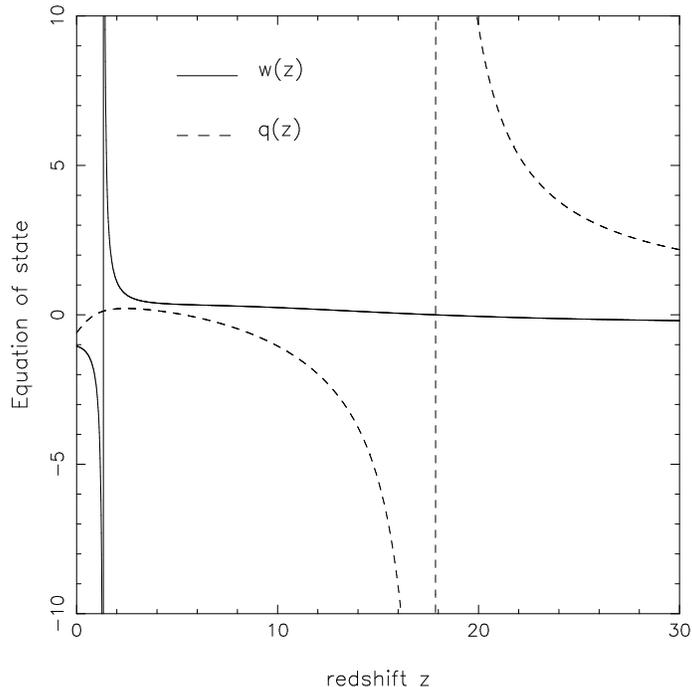} }
\bigskip
\caption{\small The effective equation of state of dark energy (solid) and the
deceleration parameter (dashed) are shown for a universe which loitered at $z
\simeq 18$. Note that the effective equation of state of dark energy becomes
infinite at low redshifts when $\Omega_{\rm m}(z) = 1$. However, this behaviour
is not reflected in the deceleration parameter, which becomes large only near
the loitering redshift. } \label{fig:state}
\end{figure*}

\item Finally, we draw attention to the fact that a loitering epoch at $z_{\rm
loit}$ can significantly alter the reionization properties of the universe at
lower redshifts. This is likely to be relevant for the following reason. One of
the main surprises emerging from the WMAP experiment was that the optical depth
to reionization was $\tau = 0.17 \pm 0.06$ \cite{wmap}. Within the framework of
`concordance cosmology' (i.e., LCDM), assuming instantaneous reionization, this
translates into a rather early epoch for the reionization redshift $z_{\rm
reion} \simeq 17 \pm 5$. (Models with `multiple reionization' epochs usually
push the reionization redshift to still higher values $z_{\rm reion} \simeq 20$
\cite{kogut03}.) `Concordance cosmology' is clearly under some pressure to
explain how the universe could have reionized at such an early cosmological
time (see \cite{reionization} for a discussion of these issues and
\cite{schwarz04} for an alternative interpretation of the large-angle WMAP
results).

Loitering has the capacity to alter these conclusions dramatically\,!
The electron scattering optical depth to a redshift $z_{\rm reion}$ is given by
\cite{peebles,dodelson} \beq
\tau(z_{\rm reion}) = c\int_0^{z_{\rm reion}}\frac{n_e(z)\sigma_T ~dz}{(1+z)H(z)}
\eeq
where $n_e$ is the electron density and $\sigma_T$ is the Thompson cross-section
describing scattering between electrons and CMB photons.
Clearly, were $H(z)$ to drop below its value in LCDM it would imply a
lower value for $z_{\rm reion}$.
Since this is precisely what happens in a loitering cosmology, one expects
${z_{\rm reion} \vert}_{\rm loitering} < {z_{\rm reion}\vert}_{\rm LCDM}$
if $z_{\rm loit} \lsim 20$.
As an example, consider the loitering models shown for illustrative purposes
in Fig.~\ref{fig:hubble}. Not surprisingly, the redshift of reionization drops to
$z_{\rm reion} \leq 12$ (from the LCDM value $z_{\rm reion} \simeq 17$)
for the loitering models shown in Fig.~\ref{fig:hubble}.
By decreasing
the  redshift of reionization as well as increasing the
age of the universe, the loitering braneworld
helps in alleviating the existing tension between the high-redshift universe and
dark-energy cosmology.

\end{itemize}

\section{Inflation in braneworld models with loitering}

The loitering braneworld models considered in the previous section place
certain constraint on the duration of the inflationary stage, as we are going
to show.  First, we note that, during the inflationary stage, the Hubble
parameter as a function of the scale factor can be approximated with a great
precision as follows (cf.\@ with (\ref{eq:loit1})):
\begin{equation} \label{eq:inflat}
H^2 (a) =  {\rho_{\rm i} (a) \over 3 m^2} - 2 {\sqrt{-C} \over \ell a^2} \, ,
\end{equation}
where $\rho_{\rm i} (a)$ is the energy density during inflation, which typically
changes very slowly with the scale factor $a$.  Since, on the contrary, the last
term in (\ref{eq:inflat})
changes rapidly during inflation, one can easily see that inflation
{\em should have a
beginning\/} in this model at the scale factor roughly given by the estimate
\begin{equation}
a_{\rm i}^2 \simeq {6 m^2 \sqrt{-C}  \over \ell \rho_{\rm i}} \, , \quad \mbox{or}
\quad \left({a_{\rm i} \over a_0}\right)^2 \simeq {2 \rho_0 \over \rho_{\rm i}}
\sqrt{\Omega_\ell \Omega_C} \, .
\end{equation}
Using (\ref{B}) from the appendix, one can write the following estimate for the
redshift $z_{\rm i}$ at the beginning of inflation:
\begin{equation}\label{eq:zinfl}
(1 + z_{\rm i})^2 \simeq {\rho_{\rm i} \over \rho_0 } \left[\sqrt{3}
\Omega_{\rm m} f(z_{\rm loit}) \left(1\!+\! z_{\rm loit} \right) \right]^{-1}
\, ,
\end{equation}
where the loitering redshift $z_{\rm loit}$ and the quantity
$f(z_{\rm loit})$, which quantifies the degree of loitering and takes values
in the range between zero and unity, are defined in the appendix.

To estimate the {\em total number\/} of the inflationary $e$-foldings, we consider a simple
model of inflation based on the inflaton $\phi$ with potential
$V(\phi) = \displaystyle \frac12 m_\phi^2 \phi^2$.  In this case, as can be shown, inflation
proceeds at the values of  the scalar field $\phi_{\rm i}  \simeq M_{\rm P}
\equiv \sqrt{8 \pi}\,m$ and ends
approximately at $\phi_{\rm f} \simeq M_{\rm P}^2 / \sqrt{12 \pi}$.  This leads to the
following relation between the typical energy density during inflation and at its end:
\begin{equation}
{\rho_{\rm i}  \over \rho_{\rm f}} \simeq 12 \pi \, .
\end{equation}
Then using (\ref{eq:zinfl}) and the estimate for the redshift at the end of inflation
\begin{equation}
1 + z_{\rm f} = {a_0 \over a_{\rm f}}  \simeq {T_{\rm rh} \over T_0} \simeq
\left({\rho_{\rm f} \over \Omega_{\rm r} \rho_0} \right)^{1/4}
\end{equation}
which assumes that preheating takes place instantaneously with effective
temperature $T_{\rm rh}$, we can estimate the redshift ratio
\begin{equation}
{z_{\rm i} \over z_{\rm f}} \simeq \left[ {4 \rho_{\rm i}^2 \over 27\rho_0
\rho_{\rm f}} \cdot {\Omega_{\rm r} \over \Omega_{\rm m}^2 f^2(z_{\rm loit})
(1\!+\!z_{\rm loit})^2 } \right]^{1/4} \simeq \left[ {16 \pi \rho_{\rm i} \over
9 \rho_0 } \cdot {\Omega_{\rm r} \over \Omega_{\rm m}^2 f^2(z_{\rm loit})
(1\!+\!z_{\rm loit})^2 } \right]^{1/4} \, .
\end{equation}
Here, $\Omega_{\rm r} \simeq 10^{-5}$ is the current value of the radiation
density parameter.

For our typical loitering redshift $z_{\rm loit} \approx 18$, for the degree of
loitering $f(z_{\rm loit}) \sim 1$, and for the estimate of the inflationary
energy density in agreement with the CMB fluctuations spectrum as
\cite{lyth-liddle} $\rho_{\rm i} / \rho_0 \sim 10^{112}$, this will restrict
the total number of inflationary $e$-foldings $N$ by
\begin{equation}
e^N \equiv {z_{\rm i} \over z_{\rm f}} \lsim 10^{26} \simeq e^{60} \, .
\end{equation}
It is interesting that the {\em total\/} number of inflationary $e$-foldings
in the loitering braneworld is close to the expected number of $e$-foldings
associated with horizon crossing in inflationary models \cite{lyth-liddle}.
The exact upper bound on the number of inflationary $e$-foldings depends
on a concrete model of braneworld inflation in the presence of loitering, and
we propose to study this issue in greater detail in a future work.

Returning to (\ref{eq:inflat}), we would like to draw the reader's attention
to the fact that, depending upon the form of the inflaton potential,
the evolution of the Hubble parameter at very early times could
have proceeded in
two fundamentally different and complementary ways:

\noindent
(i) If the shape of the inflaton potential $V(\phi)$ is
sufficiently flat, then, for a field rolling slowly,
$\rho_{\rm i} = \rho_\phi$ behaves like a slowly varying $\Lambda$-term.
As a result, the $1/a^2$ term is expected to dominate at early
times giving rise to a cosmological `bounce' ($H \simeq 0$)
when the two terms in (\ref{eq:inflat})
become comparable.

\noindent
(ii) Alternatively, it might well be that the potential $V(\phi)$
is not uniformly flat, but changes its form
and becomes steep for large values of $\phi$ (within the context of
chaotic inflation). In this case, the bounce will be avoided if,
for small values of $a$,
$\rho_{\rm i} (a)$ increases faster than the $1/a^2$ term in (\ref{eq:inflat}).
Such a rapid change in $\rho_{\rm i} (a)$ at early times
will be accompanied by the fast
rolling of the inflaton field
until the latter evolves to values where the potential is sufficiently flat
for inflation to commence.

Interestingly, both (i) and (ii) lead to departures from scale
invariance of the primordial
fluctuation spectrum on very large scales, and have been discussed in
\cite{piao03} and \cite{contaldi03}, respectively,
as providing a means of suppressing power on very large angular
scales in the CMB fluctuation spectrum.
In analogy with the discussion in these papers,
we expect that the present loitering
scenario too may
give rise to a smaller amplitude for scalar perturbations
on the largest scales, thereby providing better agreement with the
CMB anisotropy results obtained by COBE \cite{cobe} and WMAP \cite{wmap}.
These issues will be examined in greater detail in a companion paper.

\section{Loitering braneworld models without dark radiation}

It is reasonable to investigate whether spatially flat braneworld loitering
models can exist without dark radiation. This will be the purpose of the
present section.

Without the dark radiation, Eq.~(\ref{hubble}) becomes
\begin{equation}\label{hubble2}
H^2 (a) = {A \over a^3} + B + {2 \over \ell^2} \left[ 1 \pm \sqrt{1 + \ell^2
\left({A \over a^3} + B  - {\Lambda_{\rm b} \over 6} \right)} \right] \equiv
\zeta (a)  \, ,
\end{equation}
where, as before, the constants $A$ and $B$ are given by (\ref{ab}).

We look for the extrema of this function of $a$ and evaluate its values at the
extrema.  The first and second derivatives of this function are
\begin{equation}\label{prime}
\zeta'(a) = - {3 A \over a^4} \left( 1 \pm { 1 \over \sqrt{F (a)}} \right)
\end{equation}
and
\begin{equation}\label{double}
\zeta''(a) = {12 A \over a^5} \left( 1 \pm { 1 \over \sqrt{F (a)}} \right) \mp
\left( {3 A \over a^4} \right)^2 {1 \over 2  F^{3/2}(a) } \, ,
\end{equation}
respectively, where $F (a)$ denotes the expression under the square root in
(\ref{hubble2}).

Loitering occurs around the extremal point, i.e., the zero of (\ref{prime}). We
immediately see that this equation has only one zero, and only when the {\em
lower\/} sign is chosen (BRANE1 model), namely, at
\begin{equation}\label{extremum}
F (a) = 1 \quad \Rightarrow \quad {A \over a^3} + B = {\Lambda_{\rm b} \over 6}
\, .
\end{equation}
It then follows from (\ref{double}) that the second derivative is strictly
positive at this point; thus, we are dealing with a true minimum of $H^2$.

We need to evaluate the Hubble parameter at this point and to ensure that it is
only {\em slightly\/} greater than zero (the condition of loitering).  We have
\begin{equation}
H^2 \left(a_{\rm loit} \right) = {\Lambda_{\rm b} \over 6} \, ,
\end{equation}
where the extremal point $a_{\rm loit}$ is the solution of (\ref{extremum}).

One can see that this model requires {\em positive\/} value of the bulk
cosmological constant, hence, embedding of the brane in the five-dimensional
de~Sitter space rather than anti-de~Sitter space.  As can be seen from
(\ref{extremum}), this model requires also
\begin{equation}
B  \approx - {A \over a_{\rm loit}^3} < 0 \, ,
\end{equation}
i.e., $\sigma < 0$.

The universe in this model eventually evolves to the de~Sitter phase
\begin{equation}\label{ds}
H^2 \to H_0^2 = B + {2 \over \ell^2} \left[ 1 - \sqrt{ 1 + \ell^2 \left( B -
{\Lambda_{\rm b} \over 6} \right) } \right] \, ,
\end{equation}
which implies the following restriction (no `quiescent' singularity in the
future --- see \cite{ss02}):
\begin{equation}
\ell^2 \left( |B| + {\Lambda_{\rm b} \over 6} \right) \approx \ell^2 |B| < 1 \,
.
\end{equation}

In principle, we could have the condition $\ell^2 |B| \ll 1$, in which case
\begin{equation}\label{ds1}
H_0^2 \approx {\Lambda_{\rm b} \over 6 } \, ,
\end{equation}
i.e., the Hubble parameter would tend to approximately the same value that it
had at the loitering point $a_{\rm loit}$. Since the behaviour of $H^2$ is
monotonic after the extremum, this implies that, beginning from the loitering
point, the universe is effectively in the de~Sitter state with the Hubble
parameter given by (\ref{ds1}). Thus, there is no loitering phase as such in
this case, but the universe proceeds directly to the de~Sitter phase, which
should be expected since the BRANE1 model (\ref{hubble2}) in the limit $\ell
\to 0$ passes to the Randall--Sundrum model, which does not admit spatially
flat loitering solutions without dark radiation.

However, if the condition $\ell^2 |B| \sim 1$ is realized, then
\begin{equation}
H_0^2 \sim |B| \sim \ell^{-2} \gg {\Lambda_{\rm b} \over 6 } \, ,
\end{equation}
and the universe evolves to a much higher expansion rate after the period of
loitering.

The loitering braneworld model without dark radiation that we arrived at in
this section may be problematic from the viewpoint of braneworld theory since
it is embedded in de~Sitter, rather than anti-de~Sitter, five-dimensional
space.

\section{Conclusions}

We have demonstrated that a loitering universe is possible to construct within
the framework of braneworld models of dark energy. An important aspect of
braneworld loitering is that, in contrast to the conventional loitering
scenarios that demand a closed universe, loitering on the brane can easily
occur in a spatially flat cosmological model. A key role in making the brane
loiter is the presence of (negative) dark radiation --- a generic
five-dimensional effect associated with the projection of the bulk
gravitational degrees of freedom onto the brane. Our universe can loiter at
large redshifts ($z \gsim 6$) while accelerating at the present
epoch.\footnote{Although both the degree of loitering and
the loitering redshift are 
free parameters in our model whose values can be determined by matching to observations,
the loitering braneworld nevertheless does
not claim to resolve the `cosmic coincidence' conundrum associated with the
current value of the effective cosmological constant, which plagues most
dark-energy models and usually requires some degree of fine tuning of
cosmological parameters \cite{sahni04,ss00}.} During loitering, the value of
the Hubble parameter decreases steadily before increasing again. As a result,
the age of the loitering braneworld is larger than that of a LCDM universe at a
given redshift. This feature may help spur the formation of $\sim 10^9M_\odot$
black holes at redshifts $\gsim 6$ whose presence (within high redshift QSO's)
could be problematic for standard LCDM cosmology
\cite{richards03}.\footnote{From Fig.~\ref{fig:age}, we find that the age of a
loitering universe at $z \sim 6$ can be several times that in LCDM cosmology,
which is less than a billion years old at that redshift.} Loitering is also
expected to increase the growth rate of density inhomogeneities and could, in
principle, be used to reconcile structure formation models which predict a
lower amplitude of initial `seed' fluctuations with the observed anisotropies
in the cosmic microwave background (see \cite{dodelson} and references
therein). In addition, an early loitering phase could lower the redshift of
reionization from its currently high value of $z \simeq 17$ for the LCDM model
\cite{wmap}. Finally, we would like to draw attention to the fact that earlier
work on braneworlds has emphasized departure from the standard Friedmannian
behaviour either in the distant past ($z \gsim 10^{9}$) \cite{BDL} or else, in
the current epoch and remote future ($z \lsim 2$) \cite{DGP,DDG,SS,padilla}. In
this paper, we have shown that a braneworld can also show interesting
significant departures from the conventional behaviour at {\em intermediate\/}
redshifts $z \gsim 6$. It is meaningful to ask ourselves whether this feature
of braneworld cosmology is a unique aspect of the higher-dimensional action
(\ref{action}) or whether such properties are shared by a larger class of
modified gravity and string-inspired models. Perhaps future work will throw
light on this question.

{\em Acknowledgments\/}: It is a pleasure to thank Kandaswamy Subramanian,
Subhabrata Majumdar and
Carlo Contaldi for stimulating discussions and Alexei Starobinsky for valuable
comments on an early version of the paper.
The authors acknowledge support from the Indo-Ukrainian program of
cooperation in science and technology sponsored by the Department of Science
and Technology of India and Ministry of Education and Science of Ukraine.

\bigskip

\begin{appendix}
\section*{A note on the parameter space in loitering models} \label{appendix}

As pointed out in Sec.~\ref{loitering}, while not all loitering models pass through
a minimum of the Hubble parameter, a minimum value of
the ratio $H (z) / H_{\rm \Lambda CDM} (z)$ is generic and is exhibited by all models.
It is therefore  useful to supplement the definition of loitering given in (\ref{eq:loit})
by defining the loitering redshift $z_{\rm
loit}$ as the epoch associated with the
minimum of $H (z) / H_{\rm \Lambda CDM} (z)$
(both models are assumed to have the same value of $\Omega_{\rm m}$).
In order to quantify the degree of loitering, it is useful to introduce the function
\begin{equation}
\label{eq:f}
f(z) \equiv 1 - {H^2 (z) \over H^2_{\Lambda \rm CDM} (z) }
\end{equation}
where $0 \leq f < 1$. Small values $0 \leq f \leq 1/2$ imply
{\em weak loitering}, whereas larger values $1/2 < f < 1$
correspond to {\em strong loitering}.
It is straightforward to derive expressions for the loitering
redshift $z_{\rm loit}$ and the degree of loitering $f(z_{\rm loit})$ :
\begin{equation}\label{xextr}
(1 + z_{\rm loit})^4 \approx {3 \left( \sqrt{1 + \Omega_{\Lambda_{\rm b}} +
\Omega_C} + \sqrt{\Omega_\ell} \right)^2 \over \Omega_C} \, ,
\end{equation}
\begin{equation}\label{fextr}
f(z_{\rm loit}) \approx {2 \sqrt{\Omega_\ell} \left(\sqrt{1 +
\Omega_{\Lambda_{\rm b}} + \Omega_C} + \sqrt{\Omega_\ell} \right) \over
\Omega_{\rm m} (1 + z_{\rm loit})^3 } \, ,
\end{equation}
which are valid under the single assumption $\Omega_{\rm m} (1\!+\!z_{\rm loit})^3 \ll
\Omega_C (1\!+\!z_{\rm loit})^4$, or
\begin{equation}
\Omega_{\rm m} \ll \Omega_C^{3/4} \left( \sqrt{1 + \Omega_{\Lambda_{\rm b}} +
\Omega_C} + \sqrt{\Omega_\ell} \right)^{1/2} \, .
\end{equation}

From (\ref{xextr}) and (\ref{fextr}) one has the useful approximate conditions
\begin{equation}\label{A}
2 \sqrt{\Omega_\ell} \left(\sqrt{1 + \Omega_{\Lambda_{\rm b}} + \Omega_C} +
\sqrt{\Omega_\ell} \right) \approx \Omega_{\rm m} f(z_{\rm loit}) (1 + z_{\rm
loit})^3 \, ,
\end{equation}
\begin{equation}\label{B}
\Omega_C \Omega_\ell \approx \frac{3}{4}\Bigl[ \Omega_{\rm m}f(z_{\rm loit})
(1 + z_{\rm loit})\Bigr]^2~.
\end{equation}

\begin{figure*}
\centerline{ \psfig{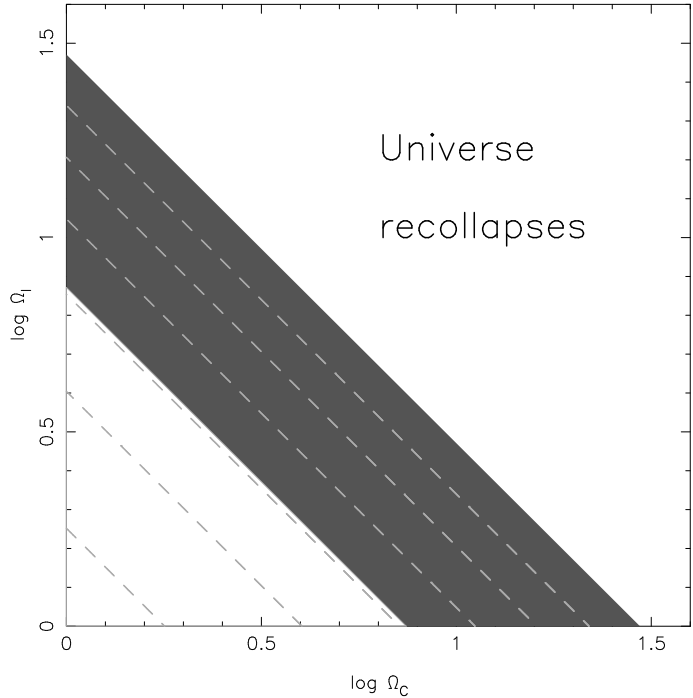} }
\bigskip
\caption{\small The parameter space $\lbrace \Omega_\ell, \Omega_C\rbrace$
is shown
for models which exhibit
(i)~weak loitering: $f (z_{\rm loit}) \leq 1/2$ in (\ref{eq:f})
(lower left corner);
(ii)~strong loitering: $1/2 < f (z_{\rm loit}) < 1$ in (\ref{eq:f})
(shaded region). The `prohibited' region corresponding to
braneworld models which recollapse {\em before\/} reaching the
present epoch is shown on the far right.
The dashed lines show contours of $\lbrace \Omega_\ell\,, \Omega_C\rbrace$
with current values of the effective equation of state:
$w_0 = -1.01, -1.015, -1.02, -1.025, -1.03, -1.035$ (from left to right).
All models loiter at
$z_{\rm loit} = 20$ and have $\Omega_{\rm m} = 0.3$.
}
\label{fig:param_space}
\end{figure*}

In practice, it is often convenient to take the values of $\Omega_{\rm m}$,
$(1\!+\!z_{\rm loit})$, and $f(z_{\rm loit})$ as control parameters and to
determine the approximate ranges of $\Omega_\ell$, $\Omega_C$, and
$\Omega_{\Lambda_{\rm b}}$ from equations (\ref{xextr})--(\ref{B}).
In Fig.~\ref{fig:param_space}, we show, as an example, the range of allowed
values for the parameter pair $\lbrace \Omega_\ell\,, \Omega_C\rbrace$ for a
model which loiters at $z_{\rm loit} = 20$ and has $\Omega_{\rm m} = 0.3$.

It is necessary to draw the reader's attention to the fact that
not every set of parameter values gives rise to a `realistic' cosmology.
For some of them, the universe recollapses before reaching the present
epoch. (The loitering braneworld shares this property
with a closed FRW universe, and the reader is referred to \cite{felten86}
for an extensive discussion of this issue.)
It is obvious that the model approaches a recollapsing
universe as the loitering parameter $ f (z_{\rm loit}) \to 1$.  Thus,
setting $f (z_{\rm loit}) = 1$ in estimate (\ref{B}), we obtain the
approximate boundary of the region of recollapsing universes
in the parameter space $\lbrace \Omega_\ell\,, \Omega_C\rbrace$:
\beq
\Omega_C \Omega_\ell \gsim \frac34 \Omega_{\rm m}^2 (1 + z_{\rm loit})^2~,
\eeq
which corresponds to the `prohibited' region in Fig.~\ref{fig:param_space}
for the particular choice of $z_{\rm loit} = 20$ and
$\Omega_{\rm m} = 0.3$.

\end{appendix}

\end{document}